\documentclass{PoS}

\newcommand{\tev}{\, {\rm TeV}}
\newcommand{\gev}{\, {\rm GeV}}

\newcommand{\be}{\begin{equation}}
\newcommand{\ee}{\end{equation}}
\newcommand{\bea}{\begin{eqnarray}}
\newcommand{\eea}{\end{eqnarray}}

\newcommand{\bi}{\begin{itemize}}
\newcommand{\ei}{\end{itemize}}

\def\kpn{K^+\rightarrow\pi^+\nu\bar\nu}

\def\klpn{K_L\rightarrow\pi^0\nu\bar\nu}

\title{FCNC Processes in the LHT Model: a 2009 Look}

\ShortTitle{FCNC in LHT: a 2009 Look}

\author{\speaker{Cecilia Tarantino}\thanks{I would like to thank the other authors of the work presented here: Monika Blanke, Andrzej J. Buras, Bj\"orn Duling and Stefan Recksiegel, for the fruitful collaboration, and the EPS organizers for the very pleasant conference realized in Krakow.}\\
        Dipartimento di Fisica, Universit\`a di Roma Tre and INFN, Sez. di Roma Tre,\\
 Via della Vasca Navale 84, I-00146 Roma, Italy\\
        E-mail: \email{tarantino@fis.uniroma3.it}}

\abstract{We present a summary of our recent analysis update of flavour changing neutral current (FCNC) processes in the Littlest Higgs model with T-parity (LHT).
The essential novelties of our update are: the removal of the logarithmic UV cutoff dependence by the inclusion of the new contribution to the $Z^0$-penguin diagrams identified by Goto et al. and by del Aguila et al., the study of the decay $K_L \to \mu^+ \mu^-$, a discussion of fine-tuning in $\varepsilon_K$ and $Br(\mu \to e \gamma)$.}

\FullConference{European Physical Society Europhysics Conference on High Energy Physics,
EPS-HEP 2009,\\
		 July 16 - 22 2009\\
		 Krakow, Poland}

\begin{document}

Little Higgs models~\cite{ArkaniHamed:2001nc} are promising candidates for New Physics (NP) being able to address the question of stability of the Higgs mass under radiative corrections.
In Little Higgs models the Higgs is naturally light as it is
identified with a Nambu-Goldstone 
boson of a spontaneously broken global
symmetry, whose gauge and Yukawa interactions are incorporated 
without generating quadratic one-loop mass corrections, through the
so-called {\it collective symmetry breaking}. 
The most economical, in matter content, Little Higgs model is the Littlest
Higgs (LH)~\cite{ArkaniHamed:2002qy}, where the global group $SU(5)$ is spontaneously broken
into $SO(5)$ at the scale $f \approx \mathcal{O}(1 \tev)$ and
the SM ew sector is embedded in an $SU(5)/SO(5)$ non-linear sigma model. 
In the LH model, the new particles appearing at the $\tev$ scales are the heavy
gauge bosons ($W^\pm_H, Z_H, A_H$), the heavy top ($T$) and the scalar triplet 
$\Phi$.
In the LH model, however, electroweak (ew) precision tests are satisfied only for quite large
values of the NP scale, $f \ge 2-3 \tev$~\cite{Han:2003wu,Csaki:2002qg}, due to
tree-level heavy gauge boson contributions and the triplet vacuum 
expectation value (vev). 
The LH model can be reconciled with ew precision tests by
introducing a discrete symmetry called T-parity~\cite{Cheng:2003ju,Cheng:2004yc}.
As T-parity explicitly forbids the tree-level contributions of  heavy gauge
bosons and the interactions that induced the triplet vev,
the compatibility with ew precision data can be obtained already for smaller 
values of the NP scale, $f \ge 500 \gev$~\cite{Hubisz:2005tx}.
Another important consequence is that particle fields are T-even or T-odd
under T-parity. The SM particles and the heavy top
$T_+$ are T-even, while the heavy gauge bosons $W_H^\pm,Z_H,A_H$ and the
scalar triplet $\Phi$ are T-odd.
Additional T-odd particles are required by T-parity: 
the odd heavy top $T_-$ and the so-called mirror fermions, i.e.,
fermions corresponding to the SM ones but with opposite T-parity and $\mathcal{O}(1 \tev)$ mass.
Mirror fermions are characterized by new flavour interactions with SM fermions
and heavy gauge bosons, which involve two new unitary 
mixing
matrices in the quark sector, $V_{Hd}$ and
$V_{Hu}$ satisfying $V_{Hu}^\dagger V_{Hd}=V_{CKM}$, and two in the lepton
sector, $V_{H\ell}$ and $V_{H\nu}$ satisfying $V_{H\nu}^\dagger V_{H\ell}=V_{PMNS}^\dagger$~\cite{Hubisz:2005bd,Blanke:2006xr}. 
Because of these new mixing matrices, the Littlest Higgs model with T-parity
(LHT) does not belong to the Minimal
Flavour Violation (MFV) class of models~\cite{Buras:2000dm,D'Ambrosio:2002ex} and significant 
effects in flavour observables are possible.
Other LHT peculiarities are the rather small number of new particles and
parameters (the SB scale $f$, the parameter $x_L$ describing $T_+$ mass and
interactions, the mirror fermion masses and $V_{Hd}$ and $V_{H\ell}$
parameters) and the
absence of new operators in addition to the SM ones.
On the other hand, one has to recall that Little Higgs models are low
energy non-linear sigma models, whose unknown UV-completion could in principle introduce for some processes a 
theoretical uncertainty reflected by a left-over logarithmic cut-off dependence.
It has been recently pointed out~\cite{Goto:2008fj,delAguila:2008zu}, however,
that in the LHT model the logarithmic UV cutoff dependence is absent at
$\mathcal{O}(v^2/f^2)$ making the predictions in this model much less sensitive
to the physics at the UV cutoff scale.
The cancellation of the logarithmic UV-cutoff dependence  in $\Delta F=1$ processes comes from an $\mathcal{O}(v^2/f^2)$ contribution to the $Z^0$-penguin diagrams overlooked in our previous analyses~\cite{Blanke:2006sb}-\cite{Blanke:2007ee} and first identified in~\cite{Goto:2008fj,delAguila:2008zu}.
We have therefore performed a new extensive analysis~~\cite{Blanke:2009am} of FCNC processes in the LHT model with the inclusion of the above mentioned $\mathcal{O}(v^2/f^2)$ contribution and by updating some theoretical and experimental inputs.
In what follows we summarize the main results obtained for both the quark and lepton sectors, underlining what is new and pointing out the channels where the largest NP effects could be seen.
For more details on the analysis we refer the reader to~\cite{Blanke:2009am}.

In the kaon system, we have studied the direct CP-violation parameter $\varepsilon_K$, which provides a stringent constraint on the LHT parameter space. We have discussed for the first time the fine-tuning required in order to satisfy the experimental constraint on $\varepsilon_K$ finding that much less fine-tuning is required in the LHT model w.r.t. the custodially protected Randall-Sundrum (RS) model~\cite{Blanke:2008zb}.
For the kaon system we also observe that large enhancements of the branching ratios $Br(\klpn)$ , $Br(\kpn)$ and $Br(K_L\to\pi^0 \ell^+\ell^-)$ with respect to the SM predictions are found to be possible.
Though the removal of the divergence reduces these enhancements by approximately a factor of two, the strong correlations among them are not modified {and provide a useful tool to distinguish the LHT model from other NP scenarios}.
Another interesting LHT correlation, which we have studied in~\cite{Blanke:2009am} for the first time, is between $Br(\kpn)$ and $Br(K_L \to \mu^+ \mu^-)$, pointing out that it is opposite and therefore distinguishable from the correlation predicted in the custodially protected RS model. Moreover, the short-distance contribution to $Br(K_L \to \mu^+ \mu^-)$  can be as large as $2.5 \cdot 10^{-9}$ in {the LHT model}, that is much larger than the SM prediction.
% \begin{figure}
% \center{\includegraphics[width=.6\textwidth]{KLmumuKp.eps}}
% \caption{\it  $Br(K_L \to 
%   \mu^+\mu^-)_{SM}$ as a function of $Br(\kpn)$. }
% \label{fig:KLmuKp}
% \end{figure}

In the $B$ system, the most interesting observable at present is the phase of $B_s^0-\bar B_s^0$ mixing. We have therefore updated the LHT prediction for the CP-asymmetry $S_{\psi \phi}$, though it is not affected by the $Z^0$-penguin contribution previously omitted. We find that $S_{\psi \phi}$ can vary in the range $[-0.4, 0.5 ]$, so that the measured deviation from the SM can be explained in the LHT model, though with some tuning of the parameters, while larger values can be easily obtained in the RS model.

In the lepton sector we find that the inclusion of the previously left-out $\mathcal{O}(v^2/f^2)$ contribution does not spoil two important results, namely the requirement of a highly hierarchical $V_{H\ell}$ matrix in order to satisfy the present upper bounds on $\mu \to e \gamma$ and $\mu^- \to e^- e^+ e^-$ and the strong correlation between the branching ratios of these two decays.
In~\cite{Blanke:2009am} we have also studied the fine-tuning required in the model in order to satisfy the present experimental bound on $Br(\mu \to e \gamma)$, finding that it is not relevant but that it would become important if the MEG experiment pushes the bound further down to $\sim10^{-13}$. 
% \begin{figure}
% \center{\includegraphics[width=0.6\textwidth]{megtun.eps}}
% \caption{\it  Fine-tuning $\Delta_{BG}(\mu\to e\gamma)$ as a function of $Br(\mu\to e\gamma)$. The shaded area represents the present (light) and future (darker) experimental constraints.}
% \label{fig:megtun}
% \end{figure}
The inclusion of the $\mathcal{O}(v^2/f^2)$ term turns out to {reduce  the LHT upper bounds for the branching ratios of the decays $\tau \to \mu(e) \pi$, $\tau \to \mu(e) \eta$, $\tau \to \mu(e) \eta'$ by approximately a factor of five}, whose enhancements with respect to the SM are nevertheless large and could be visible at a SuperB factory.
Similarly, the branching ratios for LFV $\tau$ decays with three leptons in the final state are lowered by almost an order of magnitude once the UV-cutoff dependence is cancelled, but are still large enough to be observed at a SuperB facility provided the NP scale is small enough, $f < 1 \tev$.
An important feature that is not affected by the removal of the UV-cutoff dependence is the dominance of $Z^0$-penguin and box diagram contributions relative to the dipole contributions in the decays $\ell_i^- \to \ell_j^- \ell_j^+ \ell_j^-$ and $\ell_i^- \to \ell_j^- \ell_k^+ \ell_k^-$.
This LHT feature, being in contrast to the MSSM dipole dominance, allows for a distinction between these two models, in particular when looking at ratios where the dependence on the model parameters partially cancels.
% \begin{table}
% {\renewcommand{\arraystretch}{1.5}
% \begin{center}
% \begin{tabular}{|c|c|c|c|}
% \hline
% ratio & LHT  & MSSM (dipole) & MSSM (Higgs) \\\hline\hline
% $\frac{Br(\mu^-\to e^-e^+e^-)}{Br(\mu\to e\gamma)}$  & \hspace{.8cm} 0.02\dots1\hspace{.8cm}  & $\sim6\cdot10^{-3}$ &$\sim6\cdot10^{-3}$  \\
% $\frac{Br(\tau^-\to e^-e^+e^-)}{Br(\tau\to e\gamma)}$   & 0.04\dots0.4     &$\sim1\cdot10^{-2}$ & ${\sim1\cdot10^{-2}}$\\
% $\frac{Br(\tau^-\to \mu^-\mu^+\mu^-)}{Br(\tau\to \mu\gamma)}$  &0.04\dots0.4     &$\sim2\cdot10^{-3}$ & $0.06\dots0.1$ \\\hline
% $\frac{Br(\tau^-\to e^-\mu^+\mu^-)}{Br(\tau\to e\gamma)}$  & 0.04\dots0.3     &$\sim2\cdot10^{-3}$ & $0.02\dots0.04$ \\
% $\frac{Br(\tau^-\to \mu^-e^+e^-)}{Br(\tau\to \mu\gamma)}$  & 0.04\dots0.3    &$\sim1\cdot10^{-2}$ & ${\sim1\cdot10^{-2}}$\\
% $\frac{Br(\tau^-\to e^-e^+e^-)}{Br(\tau^-\to e^-\mu^+\mu^-)}$     & 0.8\dots2.0   &$\sim5$ & 0.3\dots0.5\\
% $\frac{Br(\tau^-\to \mu^-\mu^+\mu^-)}{Br(\tau^-\to \mu^-e^+e^-)}$   & 0.7\dots1.6    &$\sim0.2$ & 5\dots10 \\\hline
% $\frac{R(\mu{Ti}\to e{Ti})}{Br(\mu\to e\gamma)}$  & $10^{-3}\dots 10^2$     & $\sim 5\cdot 10^{-3}$ & $0.08\dots0.15$ \\\hline
% \end{tabular}
% \end{center}\renewcommand{\arraystretch}{1.0}
% }
% \caption{\it  Comparison of various ratios of branching ratios in the LHT model ($f=1\tev$) and in the MSSM without \cite{Ellis:2002fe,Brignole:2004ah} and with \cite{Paradisi:2005tk,Paradisi:2006jp} significant Higgs contributions.\label{tab:ratios}}
% \end{table}

\end{document}